\documentclass[aps,prl,twocolumn,superscriptaddress,nofootinbib]{revtex4-1}

\usepackage{dcolumn}
\usepackage{gensymb} 
\usepackage{amsmath,amsfonts,amssymb,bm}
\usepackage{graphicx}			
\usepackage{graphics}			
\usepackage{xcolor}				
\usepackage[caption=false]{subfig}
\DeclareGraphicsExtensions{.pdf}

\def\EF{$E_\textrm{F}$}
\def\kF{$k_\textrm{F}$}
\def\invA{\AA$^{-1}$}

\definecolor{dkgreen}{rgb}{0.31,0.49,0.16}

\begin{document}

\title{Beyond Triplet: \\
Unconventional Superconductivity in a Spin-3/2 Topological Semimetal}

\author{Hyunsoo Kim}
\email{hyunsoo@umd.edu}
\affiliation{Center for Nanophysics and Advanced Materials, University of Maryland, College Park, MD 20742, USA}
\affiliation{Department of Physics, University of Maryland, College Park, MD 20742, USA}
\affiliation{Ames Laboratory, Department of Physics \& Astronomy, Iowa State University, Ames 50011, USA}
\author{Kefeng Wang}
\author{Yasuyuki Nakajima}
\author{Rongwei Hu}
\author{Steven Ziemak}
\author{Paul Syers}
\author{Limin Wang}
\author{Halyna Hodovanets}
\affiliation{Center for Nanophysics and Advanced Materials, University of Maryland, College Park, MD 20742, USA}
\affiliation{Department of Physics, University of Maryland, College Park, MD 20742, USA}

\author{Jonathan D. Denlinger}
\affiliation{Advanced Light Source, Lawrence Berkeley National Laboratory, Berkeley, CA 94720, USA}

\author{Philip M. R. Brydon}
\affiliation{Condensed Matter Theory Center and Joint Quantum Institute, Department of Physics, University of Maryland, College Park, MD 20742, USA}
\affiliation{Department of Physics, University of Otago, P.O. Box 56, Dunedin 9054, New Zealand}

\author{Daniel F. Agterberg}
\affiliation{Department of Physics, University of Wisconsin, Milwaukee, WI, USA}

\author{Makariy A. Tanatar}
\author{Ruslan Prozorov}
\affiliation{Ames Laboratory, Department of Physics \& Astronomy, Iowa State University, Ames 50011, USA}

\author{Johnpierre Paglione}
\email{paglione@umd.edu}
\affiliation{Center for Nanophysics and Advanced Materials, University of Maryland, College Park, MD 20742, USA}
\affiliation{Department of Physics, University of Maryland, College Park, MD 20742, USA}

\date{\today}

\begin{abstract}

In all known fermionic superfluids, Cooper pairs are composed of spin-1/2 quasi-particles that pair to form either spin-singlet or spin-triplet bound states.   
The "spin" of a Bloch electron, however, is fixed by the symmetries of the crystal and the atomic orbitals from which it is derived, and in some cases can behave as if it were a spin-3/2 particle. The superconducting state of such a system allows pairing beyond spin-triplet, with higher spin quasi-particles combining to form quintet or septet pairs. 
Here, we report evidence of unconventional superconductivity emerging from a spin-3/2 quasiparticle electronic structure in the half-Heusler semimetal YPtBi, a low-carrier density noncentrosymmetric cubic material with a high symmetry that preserves the $p$-like $j=3/2$ manifold in the Bi-based $\Gamma_8$ band in the presence of strong spin-orbit coupling. With a striking linear temperature dependence of the London penetration depth, the existence of line nodes in the superconducting order parameter $\Delta$ is directly explained by a mixed-parity Cooper pairing model with high total angular momentum, consistent with a high-spin fermionic superfluid state. 
We propose a $\mathbf{k\cdot p}$ model of the $j=3/2$ fermions to explain how a dominant $J=3$ septet pairing state is the simplest solution that naturally produces nodes in the mixed even-odd parity gap.
Together with the underlying topologically non-trivial band structure, the unconventional pairing in this system represents a truly novel form of superfluidity that has strong potential for leading the development of a new series of topological superconductors. 

\end{abstract}

\maketitle


When the spin-orbit coupling is strong enough to rearrange the order of electronic energy bands, various topological phases arise, and the interplay between superconductivity and the topologically ordered phase is of particular interest \cite{Hasan2010,Qi2011}. 
The noncentrosymmetric half-Heusler compounds containing heavy metallic elements exhibit a strong spin-orbit coupling which can invert the Bi-derived $s$-like $\Gamma_6$ and $p$-like $\Gamma_8$ bands, giving a semimetal system with non-trivial topological electronic structure \cite{Chadov2010,Lin2010,Xiao2010}. The observation of superconductivity in the $R$PtBi \cite{Goll2008,Butch2011,Tafti2013,Yan2014,Pan2013,Nakajima2015} and $R$PdBi series \cite{Nakajima2015} (where $R$=rare earth) has added a new richness to these materials that combines topological aspects of normal state band structure, superconductivity and even magnetic order \cite{Mong2010}.
In the superconducting state, non-trivial wavefunction topologies can arise both in fully gapped superconductors \cite{Fu2014} and unconventional superconductors with point or line nodes, in particular in Weyl and noncentrosymmetric superconductors \cite{Schnyder2015}.
In the latter, the lack of parity symmetry can lead to mixed even-odd parity pairing states on spin-split Fermi surfaces due to antisymmetric spin-orbit coupling \cite{Frigeri2004,Bauer2012}. 

\begin{figure} [t]
\includegraphics[width=0.9\linewidth]{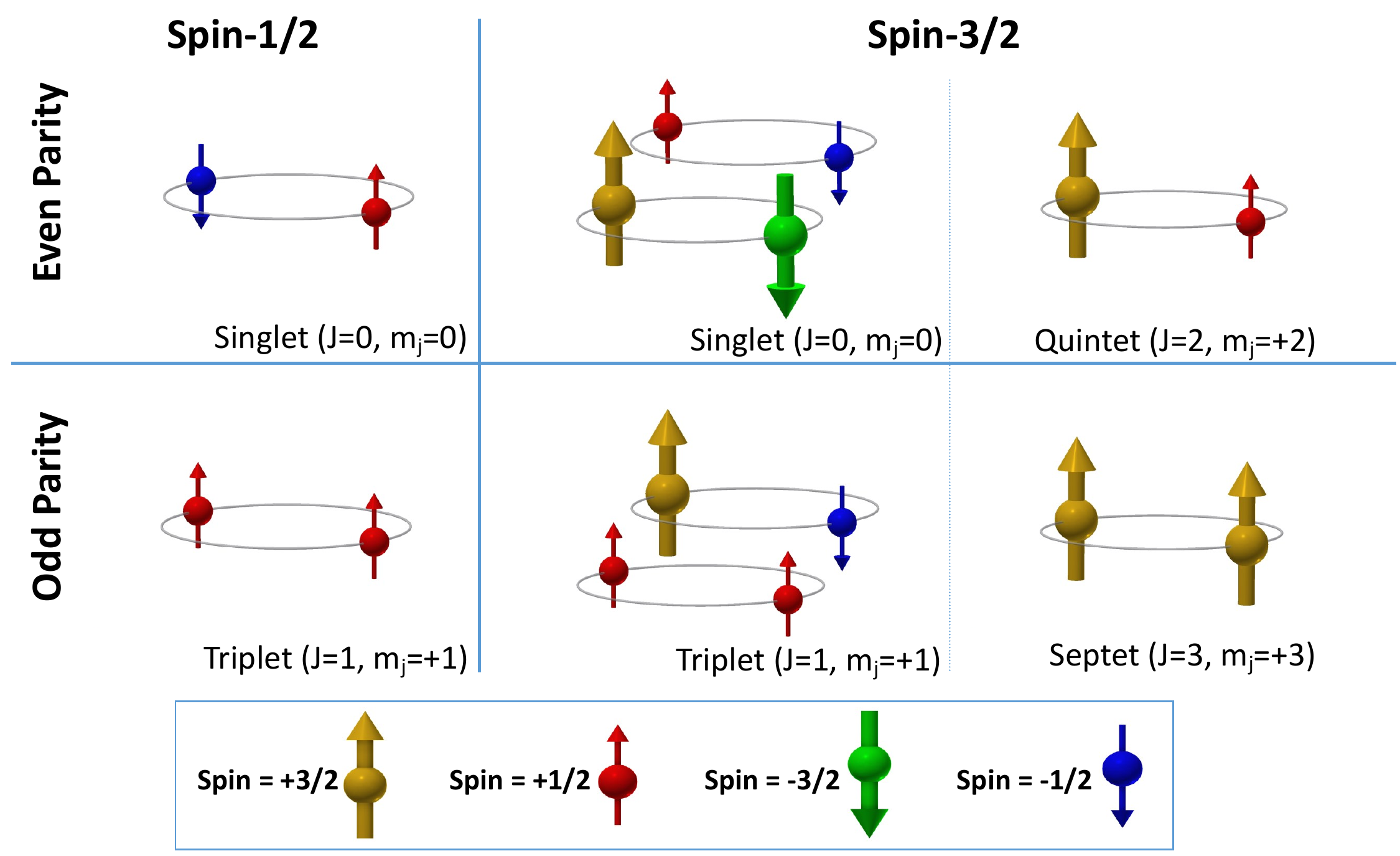}
\caption{\label{fig:pairing} {\bf High-spin Cooper pairing.} 
In conventional spin-1/2 systems, there exists four pairing channels among the spin $-1/2$ and $+1/2$ states: one spin singlet (with total angular momentum given by $J$=0 and with the $z$ component of angular momentum, $J_z$ given by $m_J$=0) and three spin-triplet, with $J$=1 and $m_J$=$-1$,0,1. The Pauli exclusion principle requires the anti-symmetric even valued $J$ states to have even spatial parity and the symmetric odd valued $J$ states to have odd spatial parity. The spin-singlet $J$=0, $m_J$=0 Cooper pair is depicted in panel (a), while the spin-triplet $J$=1, $m_J$=1 Cooper pair is depicted in panel (b). For spin-3/2 systems, there exist sixteen pairing channels among the spin $-3/2$, $-1/2$, $1/2$ and $3/2$ states: one spin-singlet with $J$=0 and $m_J$=0, three spin-triplet with $J$=1 and $m_J$=$-1$,0,1, five spin-quintet with $J$=2 and $m_J$=$-2$,$-1$,0,1,2, and seven spin-septet with $J$=3 and $m_J$=$-3$,$-2$,$-1$,0,1,2,3 (see SM for complete listing). The highest $m_J$ state for each $J$ is depicted in panels (c)-(f). In panel (c), the $J$=0, $m_J$=0 spin-singlet pair is a quantum superposition of two Cooper pairs, one pair made from spins $-1/2$ and $1/2$, and the other pair made from spins $-3/2$ and $3/2$. In panel (d), the $J$=1, $m_J$=1 spin-triplet pair is a quantum superposition of two Cooper pairs, one made from two paired spin $1/2$ states, and the other from pairing spin $3/2$ with spin $-1/2$.  Panels (e) and (f) show the $J$=2, $m_J$=2 and $J$=3, $m_J$=3 states, respectively.}
\end{figure}

The situation in half-Heusler compunds is further enriched by the $j$=3/2 total angular momentum index of the states in the $\Gamma_8$ electronic band near the chemical potential. This arises from the strong atomic spin-orbit coupling of the $s$=1/2 spin and the $l$=1 orbital angular momenta in the $p$ atomic states of Bi. The high crystal symmetry and the relatively simple band structure conspire to preserve the $j$=3/2 character of the low-energy electronic states, permitting Cooper pairs with angular momentum beyond the usual spin-singlet or spin-triplet states. In particular, as demonstrated schematically in Fig.~\ref{fig:pairing}, 
high angular momentum even- and odd-parity pairing components with quintet ($J$=2) and septet ($J$=3) states are possible through the pairing combinations of $j$=1/2 and $j$=3/2 quasiparticles, permitting a solid-state analogue of the high-spin superfluidity discussed in the context of fermionic cold atomic gases \cite{Ho1999,Yang1507}.
Such an unprecedented exotic pairing state arises from new $j$=3/2 interactions that do not appear for the spin-1/2 case, allowing new opportunities for topological superconducting states \cite{Brydon2016}.

\begin{figure} [t]
\includegraphics[width=1.0\linewidth]{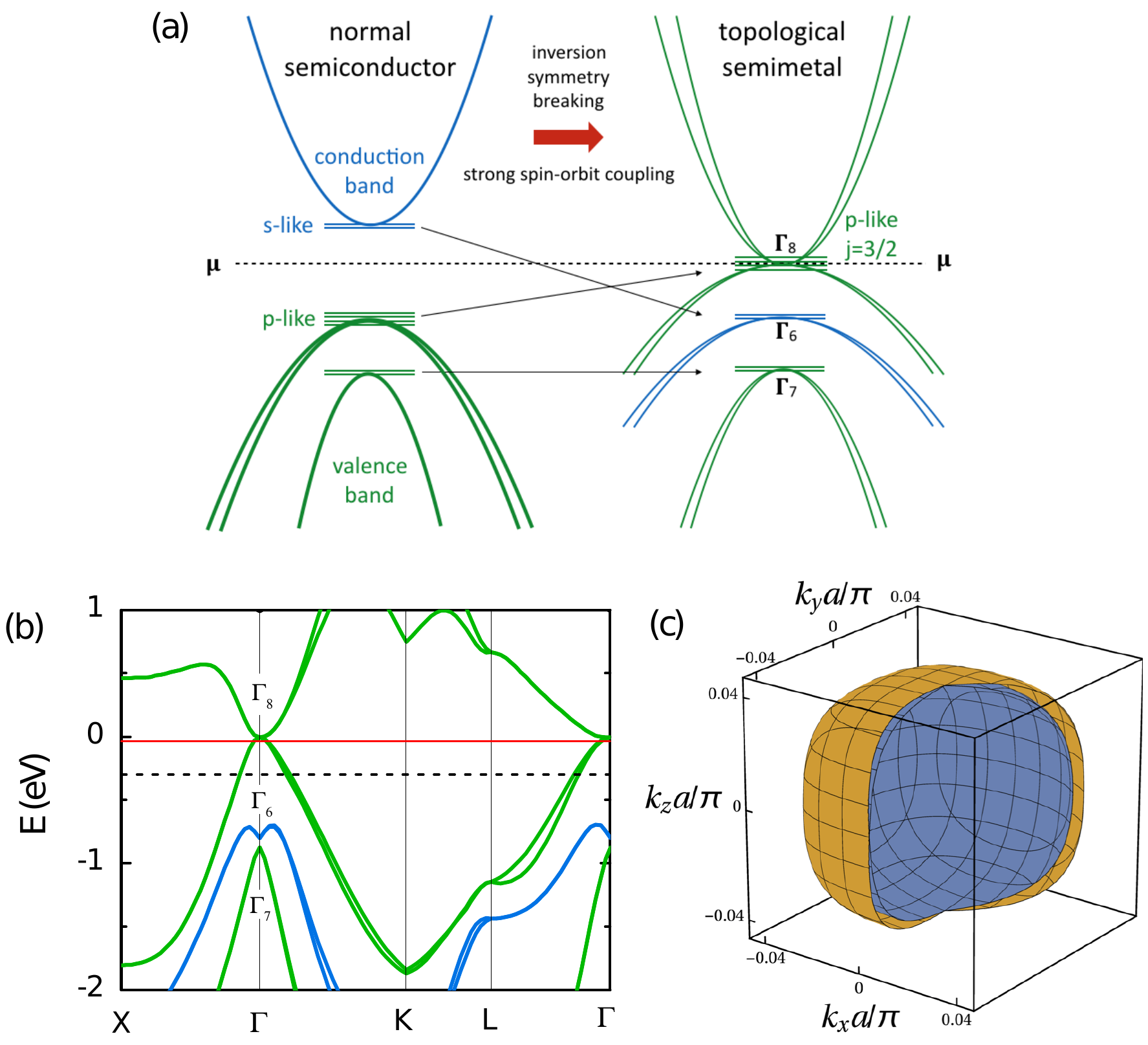}%
\caption{\label{fig:band} {\bf Topological band structure and spin-split Fermi surface of YPtBi.} 
The effect of strong spin-orbit coupling on a normal semiconductor band structure acts to produce a topological semimetal which retains $j$=3/2 character of the original $p$-like band. As shown schematically in panel (a), with increasing atomic spin-orbit coupling, the $s$-like conduction band is pushed below the chemical potential $\mu$ while the four-fold-degenerate $p$-like valence band at the $\Gamma$-point is pushed up to the chemical potential forming the $p$-like $j$=3/2 conduction band. 
The absence of parity symmetry in the YPtBi crystal structure generally causes a further splitting of the bands away from the $\Gamma$ point.
Due to the high-symmetry cubic crystal structure of YPtBi, its MBJLDA electronic structure (see text), shown in panel (b), retains the four-fold degeneracy of the $p$-like $j$=3/2 states at the $\Gamma$-point. 
The solid and dashed horizontal lines represent $\mu$ determined by quantum oscillation and angle-resolved photoemission spectroscopy experiments, respectively.
%
(c) Theoretical spin-split Fermi surfaces obtained by fitting the $j$=3/2 ${\bf k}\cdot{\bf p}$ theory in Eq. (\ref{Ham}) to the {\it ab initio} results by fixing the chemical potential at $\mu=-35$ meV (see text).} 
\end{figure}

Here we focus on the archetype topological half-Heusler YPtBi, a clean limit superconductor with an extremely small electronic density of states at the Fermi level \cite{Pagliuso1999,Butch2011}, 
corresponding to a tiny carrier density $n = 2\times10^{18}$ cm$^{-3}$ \cite{Butch2011} that rivals that of the record-holder SrTiO$_3$ \cite{Lin2013}.
The superconducting phase transition at $T_c = 0.8$ K \cite{Butch2011}
cannot be explained within the BCS theory framework, which would require a carrier density nearly three orders of magnitude larger to explain the critical temperature \cite{Meinert2016}, and the upper critical field $H_{c2}(0)$=1.5T exceeds the orbital pair-breaking limit for a conventional $s$-wave pairing state \cite{Butch2011,Bay2012}. Furthermore, the linear temperature-dependence of the upper critical field over the entire superconducting temperature range \cite{Butch2011,Bay2012} resembles that seen in the topological superconductors such as Cu$_x$Bi$_2$Se$_3$ \cite{Bay2012Bi2Se3} and Bi$_2$Se$_3$ under pressure \cite{Kirshenbaum2013}.
Furthermore, point contact spectroscopy measurements in the superconducting state exhibit several anomalous aspects (see SM for details): first, a very large energy gap of 0.2-0.3 meV, also observed in scanning tunneling spectroscopy experiments \cite{Baek2015}, corresponds to at least twice the BCS expectation of $\Delta_0$ = 1.76${k_BT_c}$ = 0.1~meV for a fully gapped superconductor with $T_c = 0.8$~K. Second, a sharp, non-Andreev-like conductance peak is observed at zero bias voltage. This stands in contrast to the typical split-peak flattened enhancement observed in $s$-wave superconductors \cite{Daghero2011}, and is very reminiscent of the odd-parity state in Sr$_2$RuO$_4$ \cite{Sengupta2002}.

\begin{figure} [t]
\includegraphics[width=1\linewidth]{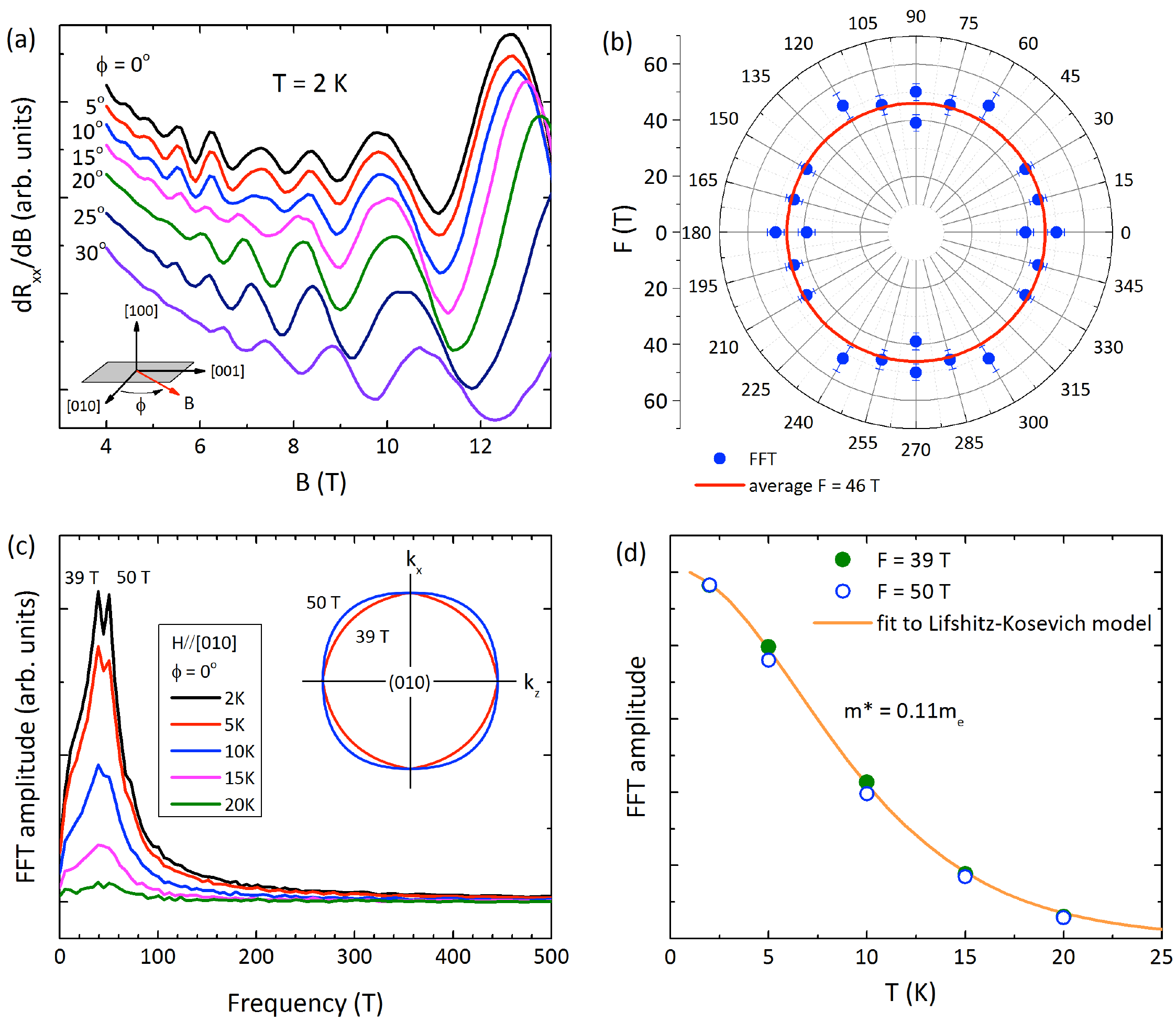}%
\caption{\label{fig:qo} {\bf Angle-dependent quantum oscillations in YPtBi.} 
Shubnikov-de Haas (SdH) oscillations were used to demonstrate the geometry of the spin-split Fermi surfaces.
(a) $dR_{xx}/dB$ vs. $B$ at various in-plane angles $\phi$ defined from crystallographic [010] direction, measured at 2~K. A node of beating oscillation is observable near 7~T for $\phi \le 10$\degree. The beating pattern changes as the angle is increased away from the high symmetry direction. Actual magnetoresistance data $R_{xx}$ are shown in SM section. 
(b) Angle-dependent frequency of SdH quantum oscillations at 2~K determined using the fast Fourier transform method on background subtracted oscillations (see SM for details). The quantum oscillation frequency is nearly independent of the field orientation (as shown by the red solid line average of 46~T, error bars from FFT frequency resolution), with two separate frequencies resolvable when the field is along the high symmetry directions, e.g., $H\parallel$ [010]. 
(c) Temperature-dependent FFT spectrum of SdH oscillations at $\phi$ = 0\degree~($H\parallel$ [010]). The two resolved frequencies of 39~T and 50~T, responsible for the beating pattern observed at $\phi \le 10$\degree, correspond respectively to the inner and outer spin-split Fermi surfaces shown in the inset.
(d) The temperature-dependent amplitude of the FFT peak is nearly identical for both spin-split Fermi surfaces and follows the Lifshitz-Kosevich expectation (solid line) with effective mass $m^*$=0.11$m_e$ for both Fermi surfaces.
}
\end{figure}


To establish a proper pairing model for YPtBi, understanding the electronic structure is essential. Generally, the crystal electric field in a fcc crystal structure splits the degenerate atomic energy levels into two-fold degenerate conduction band and two-and four-fold degenerate valence bands at the $\Gamma$-point as shown in Fig. \ref{fig:band}(a). Turning on atomic spin-orbit coupling of Bi then splits apart the two valence bands. In YPtBi, the spin-orbit coupling is sufficiently strong to invert the order of the bands, pushing the $p$-like $\Gamma_8$ band above the $s$-orbital-derived $\Gamma_6$ band. This produces a topological semimetal where the low-energy states have $j$=3/2 character.


In order to characterize the band structure in YPtBi, we employ density functional theory (DFT) calculations. The calculated bands near the Fermi level along the high symmetry points are shown in Fig. \ref{fig:band}(b), confirming the topological band inversion of $s$-like $\Gamma_6$ and $p$-like $\Gamma_8$ bands as shown previously \cite{Chadov2010,Lin2010,Xiao2010,Feng2010,Shi2015}. The absence of inversion symmetry further lifts the degeneracy of these bands. A maximum spin-splitting near the chemical potential is observed along [111] ($\Gamma$-L), whereas there is no splitting of the states along [100] ($\Gamma$-X). The theoretical chemical potential lies on the band touching point. However, experimental chemical potentials are located at $-35$ meV and $-300$ meV determined by quantum oscillation \cite{Butch2011} and ARPES \cite{Liu2016} experiments, respectively (see SM for details).



The most interesting aspect of the band inversion and the experimental position of the chemical potential in the $\Gamma_8$ band arises due to the $j$=3/2 total angular momentum, which comes from spin-orbit coupling of spin $s$=1/2 electrons in the $l$=1 $p$-orbitals of Bi.
Near the Fermi energy, we model the $\Gamma_8$ bands by a $j$=3/2 ${\bf k}\cdot {\bf p}$ theory~\cite{Dresselhaus1955}. Up to quadratic order in $k$, the single-particle Hamiltonian is 
\begin{align}
H=&\alpha k^2 +\beta\sum_i k_i^2\check{J}_i^2 {+\gamma \sum_{i\ne
j}k_ik_j\check{J}_i\check{J}_j} \notag \\&+
\delta \sum_i k_i(\check{J}_{i+1}\check{J}_i\check{J}_{i+1}-\check{J}_{i+2}\check{J}_i\check{J}_{i+2})\,,
\label{Ham}
\end{align}
where $i={x,y,z}$ and $i+1=y$ if $i=x$, {\it etc}., and $\check{J}_i$ are $4\times 4$ matrix representations of the $j$=3/2 angular momentum operators. The first line of Eq.~(\ref{Ham}) is the Luttinger-Kohn model, while the second line is the antisymmetric SOC due to the broken inversion symmetry in YPtBi. The parameters $\alpha$, $\beta$, $\gamma$, and $\delta$ are chosen by fitting to our \emph{ab initio} calculations adjusted against ARPES results, which yields $\alpha=20.5\text{ eV}a^2/\pi^2$,   $\beta=-18.5\text{ eV}a^2/\pi^2$, $\gamma=-12.7\text{ eV}a^2/\pi^2$, and $\delta = 0.06eV a/\pi$ by fixing the chemical potential at $\mu=-35$ meV, which is determined by quantum oscillations and Hall measurements \cite{Butch2011}. Here $a$ is the lattice constant taken from Ref. \cite{Haase2002}. The observed low density of hole carriers is consistent with a Fermi energy lying close to the top of the hole bands, yielding typical Fermi surfaces shown in~Fig.~\ref{fig:band}(c).

The Hamiltonian Eq.~(\ref{Ham}) has two major implications for the superconductivity in YPtBi. Firstly, since the quasiparticles in the $\Gamma_8$ band have intrinsic angular momentum $j$=3/2, they can form Cooper pairs with higher intrinsic angular momentum than allowed in the conventional theory of $j$=1/2 quasiparticle pairing; specifically, in addition to the familiar singlet ($J$=0) and triplet ($J$=1) states, we must also consider quintet ($J$=2) and septet ($J$=3) pairing (see SM for full set of states). Secondly, the absence of inversion symmetry (manifested by the antisymmetric spin-orbit coupling) implies that a stable superconducting state will be dominated by pairing between quasiparticles in time-reversed states near the Fermi energy~\cite{Frigeri2004}.  
As detailed in the Supplemental Material, this condition is generically satisfied by a mixture of conventional $s$-wave singlet pairing with an unconventional $p$-wave septet pairing state \cite{Brydon2016}. This state is discussed in more detail below and cannot occur for Cooper pairs made from pairing usual $j$=1/2 states.

To determine the spin-splitting of the true bulk Fermi surface, the angle-dependent Shubnikov-de Haas (SdH) effect was studied. First, we examine the magnetoresistance derivative, $dR_{xx}/dB$, in order to access the oscillatory component (actual magnetoresistance data $R_{xx}$ are shown in SM section). Figure~\ref{fig:qo}(a) shows $dR_{xx}/dB$ vs. $B$ at various in-plane angles $\phi$ defined from crystallographic [010] direction, measured at 2~K. A node-like feature of beating oscillations, observable near 7~T for $\phi \le 10$\degree, moves in magnetic field as the angle increases away from the highest symmetry direction. The angle-dependent SdH frequency $F$, which is proportional to the cross sectional area of Fermi surface maxima, was determined at the representative angles by employing the fast Fourier transform (FFT) technique. The symmetrized plot of $F$ vs. $\phi$ is presented in Fig.\ref{fig:qo}(b), based on actual data with $0\degree \le\phi\le 90$\degree. At first glance, the SdH frequency is almost constant around $F=46$ T (red line), which is consistent with a nearly spherical Fermi surface. A clear split-peak feature was observed when $H\parallel$ [010], i.e., $\phi=0$\degree, with two frequencies of 39 T and 50 T. The observed two frequencies correspond respectively to the inner and outer orbits of spin-split Fermi surfaces. In order to understand nature of the Fermi surfaces associated with the two frequencies, the temperature dependent amplitude of the FFT spectra was determined as shown in Fig.\ref{fig:qo}(c). As temperature rises, the split-peak feature is no longer visible above $T=10$ K due to thermal broadening of the corresponding Landau levels. A representative effective mass $m^*$=0.11$m_e$ was determined by using Lifshitz-Kosevich theory for both frequencies. The angle-dependent SdH data strongly support the theoretical spin-split Fermi surfaces.

\begin{figure} [t]
\begin{center}
\includegraphics[width=1\linewidth]{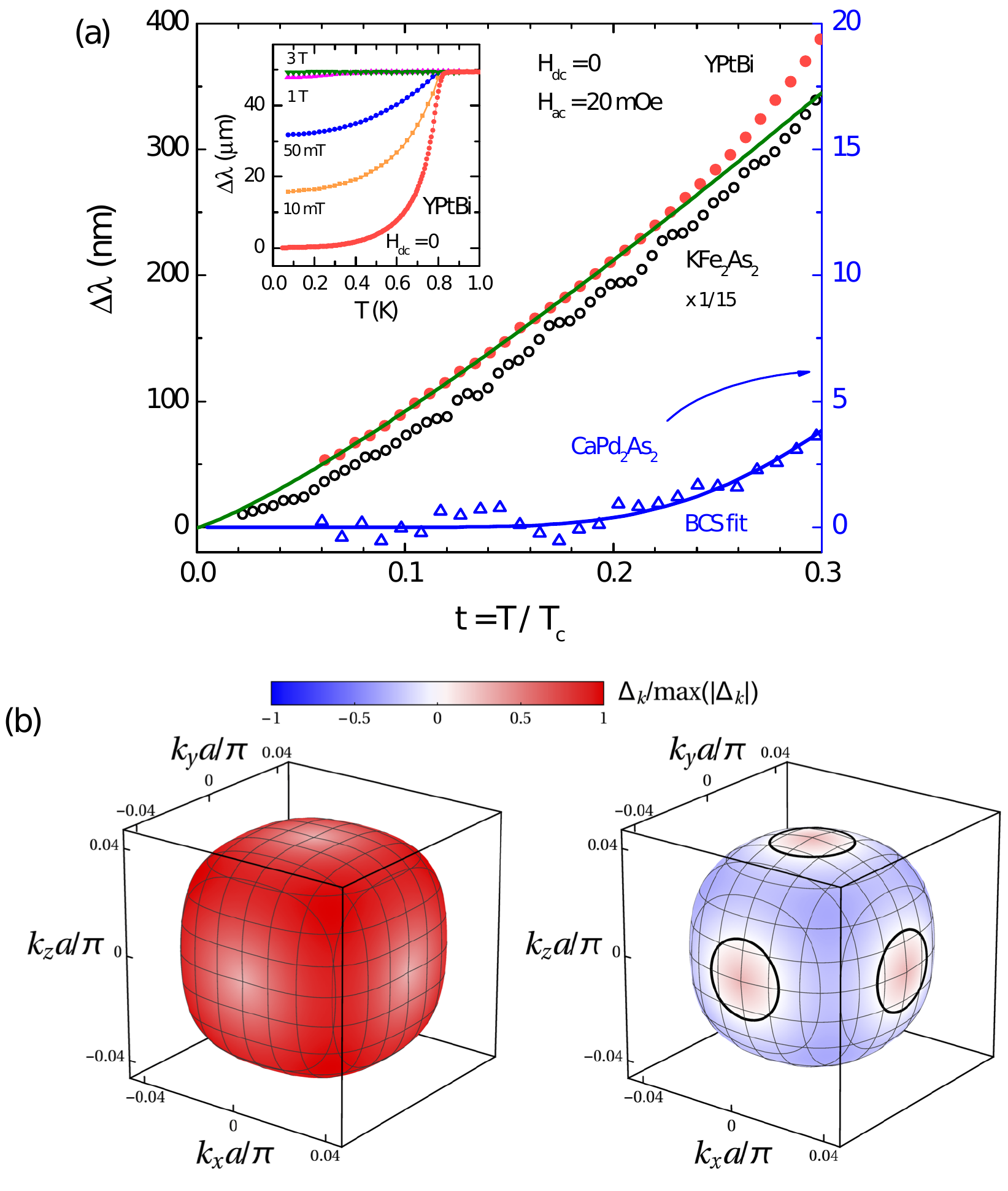}
\end{center}
\caption{\label{fig:pendepth}{\bf Evidence for line nodes and spin-3/2 singlet-septet mixed parity pairing model.} 
(a) The London penetration depth $\Delta\lambda(T)$ of YPtBi (solid circles) measured in zero dc field (20 mOe ac field) is compared in panel (a) with that of KFe$_2$As$_2$ (open circles; $T_c=3.4$ K), an unconventional superconductor with line nodes \cite{Kim2014}, and the anisotropic $s$-wave superconductor CaPd$_2$As$_2$ (open triangles; $T_c=1.3$ K) \cite{Anand2013}, both taken using the same experimental setup.
The nearly linear variation of $\Delta\lambda(T)$ is in contrast with the full-gap BCS superconductor CaPd$_2$As$_2$ data (blue line), constraining any possible pairing model to one that yields line nodes. Inset shows $\Delta\lambda$ in YPtBi under finite magnetic fields.
Due to the high intrinsic angular momentum of the $\Gamma_8$ band quasiparticles involved in Cooper pairing in YPtBi, our $j$=3/2 ${\bf k}\cdot {\bf p}$ model yields a manifold of possible pairing states with intrinsic angular momenta up to $J$=3. Incorporating the lack of inversion symmetry, the singlet-septet pairing state is the simplest even-odd parity mixture with line nodes in the pairing gap that arise from the dominant $J$=3 septet component (see text). The gap sign and magnitudes are depicted in panel (b) for the two spin-dependent Fermi surfaces of YPtBi, showing the presence of line of zero-gap nodes (black lines) situated on one of the spin-dependent surfaces.}
\end{figure}


In order to further constrain the pairing model, we focus on measurements of the temperature-dependent London penetration depth $\Delta\lambda(T)$, which is intimately related to the superconducting order parameter $\Delta$ \cite{Prozorov2011}. This approach is particularly useful in the case of YPtBi, where thermodynamic signatures of the superconducting state are difficult to measure, since $\lambda^2$ is inversely proportional to $N(0)$. In fact, the absolute value of the zero-temperature $\lambda(0)$ is 1.6~$\mu$m \cite{Bay2012}, about two orders of magnitude greater than that found in conventional superconductors with $T_c \sim 1$~K, such as zinc or aluminum.
As show in Fig. \ref{fig:pendepth}(a), a sharp transition at $T_c\approx 0.8$ K is observed to be consistent with transport measurements \cite{Butch2011,Bay2012}.

We compare the low-temperature behavior of $\Delta\lambda (T)$ in YPtBi to that of KFe$_2$As$_2$ \cite{Kim2014}, an unconventional superconductor with line nodes \cite{Kim2014}, and the anisotropic $s$-wave superconductor CaPd$_2$As$_2$ \cite{Anand2013}, both taken using an identical experimental setup. The contrast is striking, with $\Delta\lambda(T)$ in YPtBi being nearly identical to that of KFe$_2$As$_2$ and completely different from that of CaPd$_2$As$_2$. In a fully-gapped $s$-wave superconductor, the thermally activated quasiparticles are responsible for the expected exponential temperature dependence of $\Delta\lambda(T)$ at low temperatures (see SM for details), while power laws are clear signatures of nodes or zeroes in the superconducting order parameter \cite{Prozorov2011}. 
In a gap structure with line nodes, the penetration depth varies linearly with temperature at sufficiently low temperatures $(T<0.3T_c)$ in a clean sample \cite{Xu1995} (see SM for details), such as observed in the prototypical $d$-wave superconductor YBa$_2$Cu$_3$O$_{6.95}$ (YBCO) \cite{Hardy1993}, as well as the mixed-parity noncentrosymmetric superconductors CePt$_3$Si \cite{Bonalde2005} and Li$_2$Pt$_3$B \cite{Yuan2006}. 

In YPtBi, $\Delta\lambda(T)$ is best fit to a power-law function $\Delta\lambda=AT^n$ with $n=1.20\pm0.02$ and $A=1.98 \pm 0.08$ $\mu$m/K$^{1.2}$ in a temperature range that spans above $0.2T_c$. This nearly $T$-linear behavior is consistent with the expectation for a line-nodal superconductor. The observed small deviation from linearity is likely due to moderate impurity scattering, quantified by modifying the temperature dependence $\Delta\lambda(T)=bT^2/(T+T^*)$ \cite{Hirschfeld1993} with scattering rate parameter $T^*=0.07T_c$, indicating an exceptionally clean sample. The extraordinarily large power-law pre-factor $A$ in YPtBi is consistent with the London theory expectation $\lambda(0)\propto n^{-2}$ given the small carrier density of this material. 

Line nodes could in principle arise from a large number of different pairing states. However, the cubic symmetry of YPtBi imposes severe constraints on the pairing: for example, the pure $d$-wave state realized in YBCO is very unlikely here, as it would be difficult to avoid mixing with another degenerate $d$-wave state. Although symmetry permits line nodes due to an extended $s$-wave state, this requires significant fine-tuning due to the small, nearly spherical Fermi surface of the material. On the other hand, a nodal order parameter generically appears in models of YPtBi with a mixture of even- and odd-parity states, i.e. singlet pairs coexisting with triplet or septet pairs. 


Assuming an $s$-wave singlet gap, the lowest orbital angular momentum triplet ($J$=1) state is $f$-wave. Line nodes are possible for a singlet-triplet mixture, but the requirement that the triplet state have the larger gap is unlikely if quasilocal interactions give rise to superconductivity \cite{Konno1989}. Such interactions would more plausibly favor $p$-wave pairing; for the $j$=3/2 states, the noncentrosymmetric crystal symmetry permits a $p$-wave septet ($J$=3) state to generically mix with the $s$-wave singlet gap. The simplest scenario for the nodal superconductivity in YPtBi is therefore provided by a dominant septet ($J$=3) $p$-wave gap with a subdominant singlet ($J$=0) $s$-wave gap.



As shown in Fig.~\ref{fig:pendepth}(b), the gap structure resulting from this mixed singlet-septet state displays ring-shape line nodes on one of the spin-split Fermi surfaces. These line nodes are protected by a topological winding number and lead to nondegenerate surface zero-energy flat bands (see SM for details)~\cite{Schnyder2015}, which are of immense interest in the context of topological excitations. 
The mixed singlet-septet state is a natural generalization of the theory of $j$=1/2 noncentrosymmetric superconductors. 
As discussed further in the SM, however, the observation of gap line nodes and the broken inversion symmetry may be consistent with other exotic pairing states \cite{Ho1999,Boettcher2017}. Accounting for the spin-orbit coupling in Eliashberg theory reveals that odd-parity pairing may closely compete with $s$-wave singlet states \cite{Savary2017}. Nevertheless, this approach cannot reconcile the critical temperature with the very low carrier density \cite{Meinert2016}; additional pairing mechanisms, e.g. coupling to parity fluctuations \cite{Kozii2015,Wang2016}, is therefore required to properly understand the physics of the half-Heuslers.
Studying the pairing mechanism of these exotic high angular momentum pairing states, as well as their interplay with other symmetry-breaking orders \cite{Nakajima2015}, will elucidate the complexity and richness of this family of multi-faceted topological materials.


\section{Methods}

YPtBi single crystals were grown out of molten Bi with the starting composition Y:Pt:Bi = 1:1:20 (atomic ratio). The starting materials Y ingot (99.5\%), Pt powder (99.95\%) and Bi chunk (99.999\%) were put into an alumina crucible, and the crucible was sealed inside an evacuated quartz ampule. The ampule was heated slowly to 1150\degree C, kept for 10 hours, and then cooled down to 500\degree C at a 3\degree C/hour rate where the excess of molten Bi was decanted by centrifugation.

The temperature variation of London penetration depth $\Delta\lambda(T)$ was measured in a commercial dilution refrigerator by using a tunnel diode resonator (TDR) technique. The single-crystal sample with dimensions (0.29$\times$0.69$\times$0.24) mm$^3$ was mounted on a sapphire rod and inserted into a 2 mm inner diameter copper coil that produces rf excitation field with empty-resonator frequency of 22 MHz with amplitude $H_{ac} \approx 20$ mOe The shift of the resonant frequency (in cgs units), $\Delta f(T)=-G4\pi\chi(T)$, where $\chi(T)$ is the differential magnetic susceptibility, $G=f_0V_s/2V_c(1-N)$ is a constant, $N$ is the demagnetization factor, $V_s$ is the sample volume and $V_c$ is the coil volume. The constant $G$ was determined from the full frequency change by physically pulling the sample out of the coil. With the characteristic sample size, $R$, $4\pi\chi=(\lambda/R)\tanh (R/\lambda)-1$, from which $\Delta \lambda$ can be obtained \cite{Prozorov2011}.


Magnetic field-dependent magnetoresistance was determined on samples by using a standard four-probe technique. Contacts were made by using high purity silver wires and conducting epoxy, and measurements were performed in a commercial cryostat with a single-axis rotator in magnetic fields up to 14 T at temperatures as low as 2 K.

The calculated band structure of YPtBi shown in panel (a) was obtained using the WIEN2K \cite{Schwarz2002} implementation of the full potential linearized augmented plane wave method with the Tran-Blaha modified Becke-Johnson exchange-correlation potential (MBJLDA) \cite{Tran2009}, with spin-orbital coupling included in the calculation. The $k$-point mesh was taken to be 11$\times$11$\times$11, and cubic lattice constant $a=664.0(1)$~pm was obtained from Ref. \cite{Haase2002}.

\subsection{Acknowledgements}

\begin{acknowledgments}
The authors gratefully acknowledge V. Yakovenko and M. Weinert for stimulating discussion and S. Adams for her illustration works.
Research at the University of Maryland was supported by Department of Energy (DOE) Early Career Award No. DE-SC-0010605 (experimental investigations) and the Gordon and Betty Moore Foundation's EPiQS Initiative through Grant No. GBMF4419 (materials synthesis).
Work in Ames was supported by the U.S. DOE Office of Science, Basic Energy Sciences, Materials Science and Engineering Division. Ames Laboratory is operated for the U.S. DOE by Iowa State University under Contract No. DE-AC02-07CH11358.
We acknowledge support from Microsoft Station Q, LPS-CMTC, and JQI-NSF-PFC (P.M.R.B), and
the NSF via DMREF-1335215 (D.F.A.).  
ARPES experiments were supported by the U.S. DOE at the Advanced Light Source (DE-AC02-05CH11231).
\end{acknowledgments}


\section{Supplementary Materials}

\subsection{Spin-3/2 pairing}

In YPtBi, the electronic $\Gamma_8$ representation responsible for the states near the chemical potential can be described by a  $j=3/2$ basis with four basis elements: $\{|\tfrac{3}{2}\rangle,|\tfrac{1}{2}\rangle,|-\tfrac{1}{2}\rangle,|-\tfrac{1}{2}\rangle\}$. Physically this basis stems from $l=1$ $p$-states coupled to $s=1/2$ spin. In this direct product space, the basis elements can be expressed as:
\begin{eqnarray}
|\tfrac{3}{2}\rangle=&\tfrac{1}{\sqrt{2}}[-|p_x,\tfrac{1}{2}\rangle-i|p_y,\tfrac{1}{2}\rangle]\\
|\tfrac{1}{2}\rangle=&\frac{1}{\sqrt{6}}[2|p_z,\tfrac{1}{2}\rangle-|p_x,-\tfrac{1}{2}\rangle-i|p_y,-\tfrac{1}{2}\rangle]\\
|-\tfrac{1}{2}\rangle=&\frac{1}{\sqrt{6}}[2|p_z,-\tfrac{1}{2}\rangle+|p_x,\tfrac{1}{2}\rangle-i|p_y,\tfrac{1}{2}\rangle]\\
|-\tfrac{3}{2}\rangle=&\frac{1}{\sqrt{2}}[|p_x,-\tfrac{1}{2}\rangle-i|p_y,-\tfrac{1}{2}\rangle].
\end{eqnarray}
Cooper pairs can be constructed from these $\Gamma_8$ states. In particular using the angular momentum addition rule: $\tfrac{3}{2}\otimes \tfrac{3}{2}=3\oplus2\oplus1\oplus0$, we can classify the sixteen possible Cooper pairs as follows:
\begin{widetext}
\noindent $J=0$ singlet state
\begin{equation}
|J=0,m_J=0\rangle=\tfrac{1}{2}\Big (|\tfrac{3}{2},-\tfrac{3}{2}\rangle-|-\tfrac{3}{2},\tfrac{3}{2}\rangle-|\tfrac{1}{2},-\tfrac{1}{2}\rangle+|-\tfrac{1}{2},\tfrac{1}{2}\rangle \Big );
\end{equation}
$J=1$ triplet states
\begin{eqnarray}
|J=1,m_J=1\rangle=& \frac{1}{\sqrt{10}}\Big (\sqrt{3}|\tfrac{3}{2},-\tfrac{1}{2}\rangle-2|\tfrac{1}{2},\tfrac{1}{2}\rangle+\sqrt{3}|-\tfrac{1}{2},\tfrac{3}{2}\rangle\Big ) \nonumber \\
|J=1,m_J=0\rangle=& \frac{1}{\sqrt{20}}\Big (3|\tfrac{3}{2},-\tfrac{3}{2}\rangle-|\tfrac{1}{2},-\tfrac{1}{2}\rangle-|-\tfrac{1}{2},\tfrac{1}{2}\rangle+3|-\tfrac{3}{2},\tfrac{3}{2}\rangle \Big ) \nonumber \\
|J=1,m_J=-1\rangle=& \frac{1}{\sqrt{10}}\Big (\sqrt{3}|-\tfrac{3}{2},\tfrac{1}{2}\rangle-2|-\tfrac{1}{2},-\tfrac{1}{2}\rangle+\sqrt{3}|\tfrac{1}{2},-\tfrac{3}{2}\rangle\Big );
\end{eqnarray}
$J=2$ quintet states
\begin{eqnarray}
|J=2,m_J=2\rangle=& \frac{1}{\sqrt{2}}\Big (| \tfrac{3}{2},\tfrac{1}{2}\rangle-|\tfrac{1}{2}, \tfrac{3}{2}\rangle \Big ) \nonumber \\
|J=2,m_J=1\rangle=& \frac{1}{\sqrt{2}}\Big (| \tfrac{3}{2},-\tfrac{1}{2}\rangle-|-\tfrac{1}{2}, \tfrac{3}{2}\rangle \Big  )\nonumber \\
|J=2,m_J=0\rangle=& \frac{1}{2}\Big (| \tfrac{3}{2},- \tfrac{3}{2}\rangle+|\tfrac{1}{2},-\tfrac{1}{2}\rangle-|\tfrac{1}{2},-\tfrac{1}{2}\rangle-|- \tfrac{3}{2}, \tfrac{3}{2}\rangle \Big ) \nonumber \\
|J=2,m_J=-1\rangle=& \frac{1}{\sqrt{2}}\Big (|- \tfrac{3}{2},\tfrac{1}{2}\rangle-|\tfrac{1}{2},- \tfrac{3}{2}\rangle \Big  )\nonumber \\
|J=2,m_J=-2\rangle=& \frac{1}{\sqrt{2}}\Big (|- \tfrac{3}{2},-\tfrac{1}{2}\rangle-|-\tfrac{1}{2},- \tfrac{3}{2}\rangle \Big );
\end{eqnarray}
and $J=3$ septet states
\begin{eqnarray}
|J=3,m_J=3\rangle=& |\tfrac{3}{2},\tfrac{3}{2}\rangle \nonumber \\
|J=3,m_J=2\rangle=& \frac{1}{\sqrt{2}}\Big (|\tfrac{3}{2},\tfrac{1}{2}\rangle+|\tfrac{1}{2},\tfrac{3}{2}\rangle \Big )  \nonumber \\
|J=3,m_J=1\rangle=& \frac{1}{\sqrt{5}}\Big (|\tfrac{3}{2},-\tfrac{1}{2}\rangle+\sqrt{3}|\tfrac{1}{2},\tfrac{1}{2}\rangle+|-\tfrac{1}{2},\tfrac{3}{2}\rangle\Big ) \nonumber \\
|J=3,m_J=0\rangle=& \frac{1}{\sqrt{20}}\Big (|\tfrac{3}{2},-\tfrac{3}{2}\rangle+3|\tfrac{1}{2},-\tfrac{1}{2}\rangle+3|-\tfrac{1}{2},\tfrac{1}{2}\rangle+|-\tfrac{3}{2},\tfrac{3}{2}\rangle \Big ) \nonumber \\
|J=3,m_J=-1\rangle=& \frac{1}{\sqrt{5}}\Big (|\tfrac{3}{2},-\tfrac{1}{2}\rangle+\sqrt{3}|-\tfrac{1}{2},-\tfrac{1}{2}\rangle+|\tfrac{1}{2},-\tfrac{3}{2}\rangle\Big ) \nonumber \\
|J=3,m_J=-2\rangle=& \frac{1}{\sqrt{2}}\Big (|-\tfrac{3}{2},-\tfrac{1}{2}\rangle+|-\tfrac{1}{2},-\tfrac{3}{2}\rangle \Big ) \nonumber \\
|J=3,m_J=-3\rangle=&|-\tfrac{3}{2},-\tfrac{3}{2}\rangle.
\end{eqnarray}
\end{widetext}

\subsection{Band structure and $\mathbf{k\cdot p}$ model}

\begin{figure}
\includegraphics[width=0.9\linewidth]{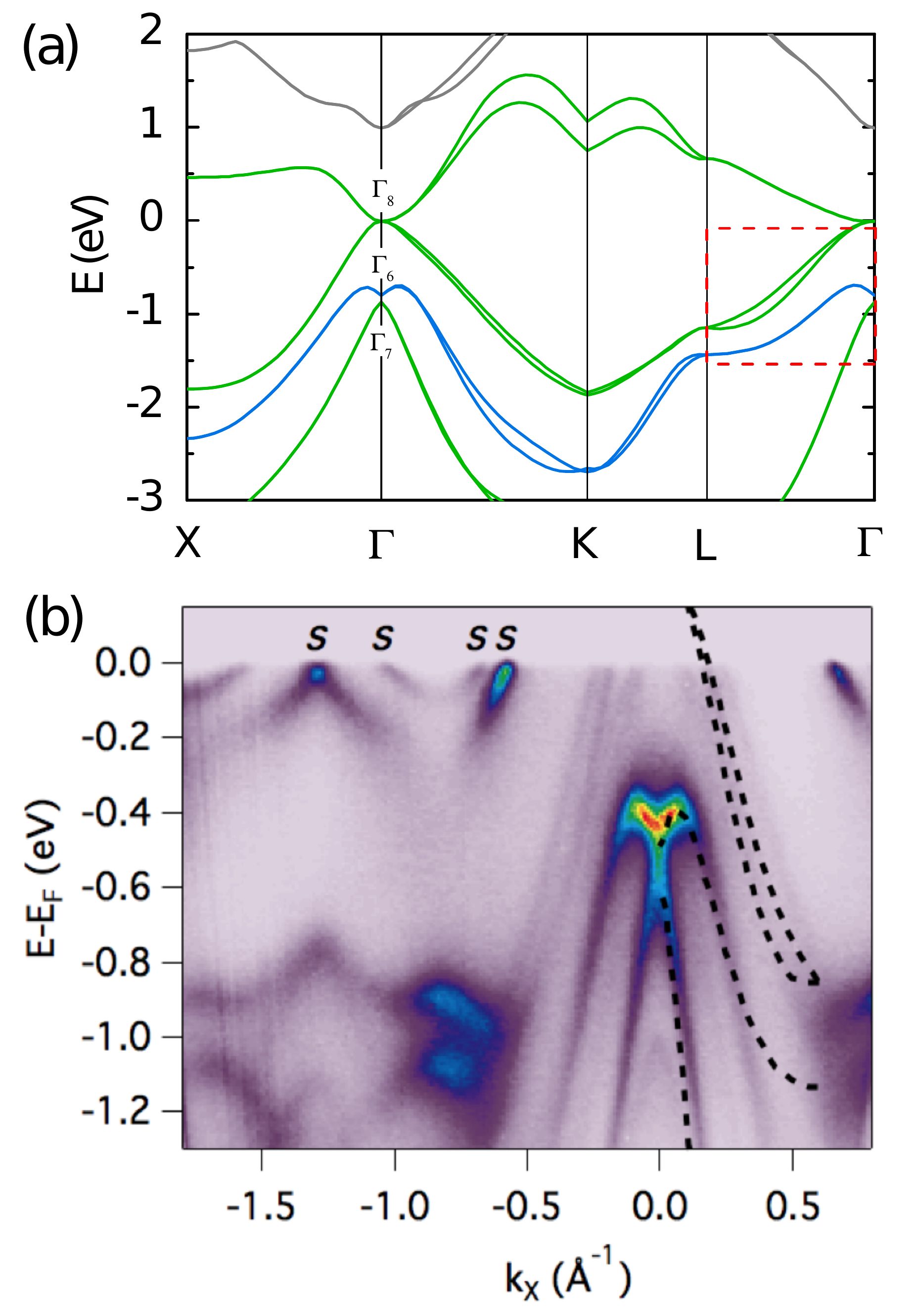}%
\caption{\label{fig:bandsi} (a) Calculated band structure of YPtBi obtained using the full potential linearized augmented plane wave method with the Tran-Blaha modified Becke-Johnson exchange-correlation potential (MBJLDA). (b) Results of angle-dependent photoemission spectroscopy (ARPES) measurement done on Bi-terminated (111) surface. The black dashed lines represent the calculated band structure along $\Gamma$-L. The chemical potential in ARPES result is about 0.32 eV below that of theoretical result.} 
\end{figure}

\subsubsection{Band structure}
Figure~\ref{fig:bandsi} presents the band structure of YPtBi. The calculated band structure of YPtBi shown in panel (a) was obtained using the WIEN2K \cite{Schwarz2002} implementation of the full potential linearized augmented plane wave method with the Tran-Blaha modified Becke-Johnson exchange-correlation potential (MBJLDA) \cite{Tran2009}, with spin-orbital coupling included in the calculation. The $k$-point mesh was taken to be 11$\times$11$\times$11, and cubic lattice constant $a=664.0(1)$~pm was obtained from Ref. \cite{Haase2002}. The $s$-like band $\Gamma_6$ lies below the $p$-like band $\Gamma_8$, reflecting the nontrivial topology, consistent with previous calculations \cite{Shi2015,Feng2010}. Note the splittings will disappear in a non-relativistic calculation, and also it is common to have these spin splittings vanish at high symmetry points and along high symmetry directions. This is because there are symmetry elements that require band degeneracies, which are removed when we move away from these special points. This splitting evidences a $j$=3/2 analogue of antisymmetric spin-orbit coupling due to broken inversion symmetry, similar to Rashba spin-orbit coupling in tetragonal systems \cite{Dresselhaus1955}.

Figure~\ref{fig:bandsi}(b) shows ARPES results done on Bi-terminated (111) surface. Numerous bands are observed to cross the Fermi-level (\EF) from the zone boundary to normal emission. Consistent with ARPES measurements of LuPtBi and GdPtBi \cite{Liu2011}, the photon energy dependence of most of these bands follows strong vertical streaks along $k_z$ indicative of 2D surface states.  Theoretical slab calculations \cite{Liu2011}  have determined that the surface states, labeled $s$ in Fig. \ref{fig:bandsi}(b), originate from a Bi-terminated (111) cleavage plane, and that they are of non-topological origin owing to an even number of \EF-crossings with the surface Brillouin zone. Aging of the surface in a poorer storage vacuum for a week is observed to suppress all the surface states and leave a single fuzzy broad hole-band feature that reaches to \EF.   A similar inner-hole band pocket was also observed as a single-band with weak intensity in LuPtBi and GdPtBi \cite{Liu2011} and assumed to be a pair of nearly degenerate bands from comparison to their slab calculations. 

A characteristic Rashba-like splitting of two hole bands is shown in Fig. \ref{fig:bandsi}(b) at $\approx$ 0.5 eV binding energy which also appear in bulk band structure calculations, but at an energy of $\approx$ -0.8 eV below \EF.  Shifting the theory $\Gamma$-L bands to higher energy by $\approx$ 0.3 eV to align to the Rashba-like split bulk bands, as shown in overplotted dashed lines, causes two other nearly-degenerate hole-like bulk bands that originally just touch \EF~at a semimetal point, to form a hole-pocket. This large 0.3 eV chemical potential shift in the ARPES measurement, relative to the theory calculation, reflects a possible charge imbalance at the cleaved surface and resultant band bending relative to the bulk.  

The experimental quantum oscillation frequency of $F=45$ T \cite{Butch2011} corresponds to a cross sectional area of a hypothetical spherical Fermi surface $A_F\approx 0.43$ nm$^{-2}$ and a \kF$\approx$0.037 \invA, using the Onsager relation $F=\phi_0 A_F/2\pi^2$ where $A_F=\pi k_F^2$ and $\phi_0=2.07\times 10^{-7}$ G cm$^{-2}$, which is much smaller than the ARPES \kF\ values in Fig.~\ref{fig:bandsi}(b) of $\approx$ 0.1 \invA.  This supports the scenario of a charge imbalance at the surface and band bending with the bulk chemical potential only $\approx$35 meV below the hole-band maximum. 

\subsubsection{$\mathbf{k\cdot p}$ model}

Due to the presence of time-reversal and inversion symmetry for $\delta=0$, the eigenstates can be labeled by a pseudospin-1/2 index (we emphasize that the pseudospin-1/2 cannot be chosen to transform like a true spin-1/2 under the symmetries of the crystal, however, due to the spin-3/2 nature of the electron states).

We treat the antisymmetric spin-orbit coupling (ASOC) as a perturbation of the Luttinger-Kohn
model, which has doubly-degenerate eigenenergies 
\begin{equation}
\epsilon_{\bf k,\pm}=\left(\alpha
+{ \frac{5}{4}}\beta\right)|{\bf k}|^2\pm\beta\sqrt{\sum_{i}\left[k_i^4+\left(\tfrac{3\gamma^2}{\beta^2}-1\right)k_i^2k_{i+1}^2\right]}. \label{eq:centrodisp}
\end{equation}
Due to the presence of time-reversal and inversion symmetry for
$\delta=0$, the 
eigenstates can be labelled by a pseudospin-$1/2$ index. Proceeding
via degenerate perturbation theory, we now include the ASOC by projecting
it into the pseudospin basis for each band, hence obtaining two effective
pseudospin-$1/2$ Hamiltonians 
\begin{equation}
H_{\rm{eff},\pm}=\epsilon_{\bf
k,\pm}\hat{s}_0 +{\bf g}_{{\bf k},\pm}\cdot \hat{\bf s} \label{eq:effHam}
\end{equation}
where $\hat{s}_\mu$ are the
Pauli matrices for the pseudospin, and the vector ${\bf g}_{{\bf
k},\pm} = -{\bf g}_{-{\bf k},\pm}$ represents the effective ASOC in the
pseudospin-$1/2$ basis of the band $\epsilon_{{\bf k},\pm}$. The
expression for ${\bf g}_{{\bf k},\pm}$ is complicated and
depends upon the choice of pseudospin basis; an analytic expression
for $|{\bf g}_{{\bf k},\pm}|$ is given elsewhere~\cite{Brydon2016}. The effective 
pseudospin-$1/2$ Hamiltonians can be 
diagonalized by going over to the helicity basis, yielding dispersions
$E_{{\bf k},\eta=\pm,\nu=\pm} = \epsilon_{{\bf k},\eta} + \nu|{\bf 
g}_{{\bf k},\eta}|$,
where the values of $\eta$ and $\nu$ are independent of one
another. In particular, the two spin-split Fermi surfaces are labelled
by opposite values of $\nu$. This approximation is in excellent
agreement with the exact solutions of the ${\bf k}\cdot{\bf p}$ for
small antisymmetric SOC.

In weak-coupling Bogoliubov-de Gennes theory the pairing is modelled
by a term in the Hamiltonian of the form
\begin{equation}
H_{\text{pair}}=\sum_{\bf
k}\sum_{j,j'=-3/2}^{3/2}\Big\{\Delta_{j,j'}({\bf k})
c^{\dagger}_{{\bf k},j}c^{\dagger}_{-{\bf k},j'}+\text{H.c.}\Big\}.
\end{equation}
We restrict our attention to gaps in the $A_1$ irreducible
representation of the $T_d$ point group (i.e. with the full symmetry
of the lattice). Allowing pairing in at most a relative $p$-wave
(assuming quasi-local interactions are responsible for the
superconductivity, higher-order momentum-dependence is unlikely~\cite{Konno1989}), we
have the general gap
\begin{widetext}
\begin{equation}
\check{\Delta}({\bf k})=\Delta_s \left(\begin{array}{cccc}
0 & 0 & 0 & 1 \\
0 & 0 & -1 & 0 \\
0 & 1 & 0 & 0 \\
-1 & 0 & 0 & 0
\end{array}\right) + \Delta_p\left(
                  \begin{array}{cccc}
                    \frac{3}{4}k_- & \frac{\sqrt{3}}{2}k_z &
                  \frac{\sqrt{3}}{4}k_+ & 0 \\
                    \frac{\sqrt{3}}{2}k_z & \frac{3}{4}k_+ & 0 & -\frac{\sqrt{3}}{4}k_- \\
                    \frac{\sqrt{3}}{4}k_+ & 0 & -\frac{3}{4}k_- & \frac{\sqrt{3}}{2}k_z \\
                    0 & -\frac{\sqrt{3}}{4}k_- & \frac{\sqrt{3}}{2}k_z & -\frac{3}{4}k_+ \\
                  \end{array}
                \right) \label{eq:septet}
\end{equation}
\end{widetext}
where $k_{\pm}=k_x\pm i k_y$. This constitutes a mixed state involving
$s$-wave singlet pairing with strength $\Delta_s$ and $p$-wave {\it septet}
pairing with strength $\Delta_p$. The gap near the Fermi energy can be
found by projecting~Eq.~(\ref{eq:septet}) into the effective
pseudospin-$1/2$ bands, yielding
\begin{equation}
\Delta_{\text{eff},\pm}=\left[\Delta_s + (\Delta_p/\delta)({\bf g}_{{\bf
k},\pm}\cdot\hat{\bf s})\right]i\hat{s}_y\,.
\end{equation}
This describes a mixture of pseuodospin-singlet and
pseudospin-triplet pairing. Importantly, the ${\bf
d}$-vector of the effective pseudospin-triplet pairing is parallel to
the ASOC vector ${\bf g}_{{\bf k},\pm}$. As pointed out by Frigeri {\it et
al.}~\cite{Frigeri2004}, this alignment makes the pseudospin-triplet component
immune to the pair-breaking effect of the ASOC; for sufficiently large 
ASOC, it is the only stable odd-parity gap. If the 
singlet state is subdominant, the resulting gap displays line nodes on
one of the spin-split Fermi surfaces. 

\begin{figure}
\includegraphics[width=1\linewidth]{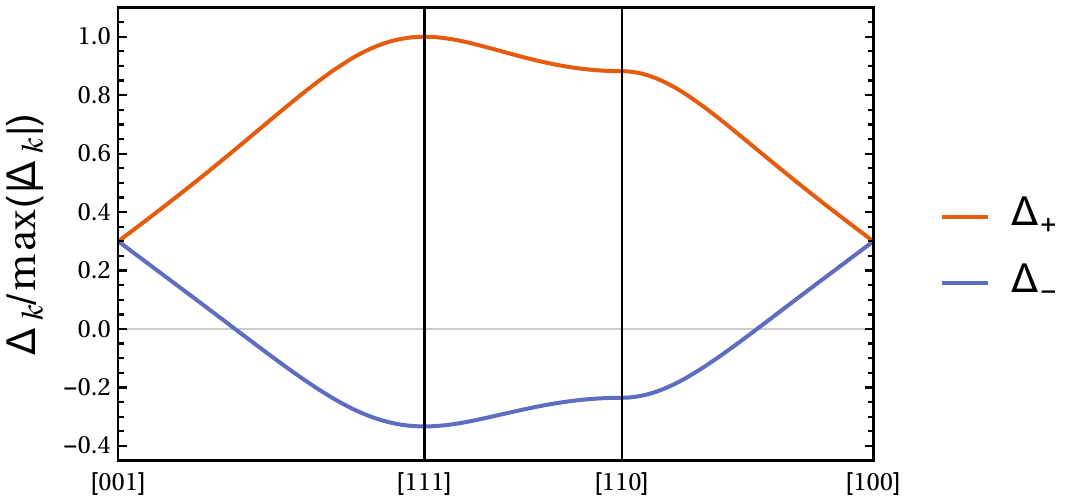}%
\caption{\label{fig:gapdir} Normalized magnitude of each gap $\Delta_k/\textmd{max}(|\Delta_k|)$ along high symmetry points are presented for full gap and nodal gap in red and blue, respectively.
Whereas the fully gapped branch (red) contributes only thermally activated quasiparticles at low temperatures, the line-node branch (blue) manifests a linear temperature variation of the London penetration depth $\Delta\lambda\propto T$, consistent with experimental observations in YPtBi.} 
\end{figure}

Figure \ref{fig:gapdir} shows distribution of gap amplitudes $\Delta_k/\textmd{max}(|\Delta_k|)$ along high symmetry points on the spin-split Fermi surfaces of YPtBi, with a full gap on the outer Fermi surface and gap with line nodes on the inner Fermi surface (see Fig. 4(b) in the main text). Whereas the fully gapped branch (red) contributes only thermally activated quasiparticles at low temperatures, the line-node branch (blue) manifests a linear temperature variation of the London penetration depth $\Delta\lambda\propto T$, consistent with experimental observations in YPtBi.

Although the nodal gap prevents us from defining a global topological
invariant, the nodal lines themselves represent a nontrivial
topological defect in the Brillouin zone. Specifically, the
Bogoliubov-de Gennes Hamiltonian $H_{BdG}({\bf k})$ belongs to
Altland-Zirnbauer class 
DIII, which implies that it can be brought into off-diagonal
form. This allows us to define the winding number
\begin{equation}
W_{\cal L} = \frac{1}{2\pi}\text{Im}\oint_{\cal
L}dl \text{Tr}\left\{\nabla_{l}\ln(D_{\bf k})\right\}
\end{equation}
where $D_{\bf k}$ is the upper off-diagonal block of the Hamiltonian~\cite{Schnyder2011,Brydon2011,Schnyder2015}.
The winding number $W_{\cal L}$ takes an integer value along any
closed path ${\cal L}$ in the Brillouin zone that does not intersect a
gap node. Moreover, it is only nonzero if the path ${\cal L}$
encircles a line node, defining the topological charge of the
node which in our case evaluates to $\pm1$. This topological charge
ensures the existence of a nondegenerate 
zero-energy surface flat band within the projection of the line node
in the surface Brillouin zone~\cite{Schnyder2011,Brydon2011,Schnyder2015}. 

Some alternatives to the mixed singlet-septet state proposed here
should be noted. In particular, for purely local pairing interactions,
there are five additional $s$-wave states with quintet total angular
momentum~\cite{Brydon2016}. Within a weak-coupling theory, these
combine to give time-reversal symmetry-breaking states gaps with Weyl
point nodes, and in some cases also line nodes. While such pairing
states also topological, they do not yield robust zero-energy surface
flat bands~\cite{Kobayashi2015}.

\subsection{London penetration depth}

The temperature variation of London penetration depth is intimately related to the superconducting order parameter $\Delta$. Within a weak coupling Eilenberger quasiclassical formulation with the perturbation theory of a weak magnetic field \cite{Prozorov2011},
\begin{equation}\label{eq:lambda}
(\lambda^2)^{-1}_{ik} = \frac{16\pi^2 e^2 T}{c^2}N(0)\sum_\omega\left< \frac{\Delta^2 v_i v_k}{(\Delta^2+\hbar^2\omega^2)^{3/2}}\right>
\end{equation}
where $N(0)$ is the total density of states at Fermi level per spin, $v$ is the Fermi velocity, and $\omega$ is the Matsubara frequency. 
Measurement of the London penetration depth utilizes a small excitation field $H_{ac}<2\mu$T, so the Eq. (\ref{eq:lambda}) is valid in absence of $H_{dc}$. 

In an $s$-wave superconductor, an exponential behavior of $\Delta\lambda(T)/\lambda(0)=\sqrt{\pi \Delta_0/2k_BT}\exp(-\Delta_0/k_BT)$ can be deduced from Eq. (\ref{eq:lambda}) for a constant gap $\Delta=\Delta_0$, while 
in a $d$-wave superconductor $\Delta\lambda(T)$ varies linearly with temperature as $\Delta\lambda(T)=\frac{2 \lambda(0) \ln2}{\mu\Delta_0}T$, at sufficiently low temperatures in a clean sample \cite{Xu1995}. 
Here, $\mu$ is the angular slope parameter near the node, {\it e.g.,} $\mu=2$ for a $d$-wave gap $\Delta=\Delta_0(k_x^2-k_y^2)$. 
To compare the experimental result for YPtBi to the $d$-wave gap expectation, one can fix the temperature power of $\Delta\lambda(T)$ to $n=1$ and obtain a slope prefactor $A=1.5$ $\mu$m/K from fitting, yielding $2\Delta_0/k_B T_c\approx 4/\mu$. Fixing $\mu=2$ for the $d$-wave case gives $2\Delta_0\approx 2 k_B T_c$, which is slightly smaller than the weak-coupling BCS value of $2\Delta_0 = 3.52k_B T_c$, and significantly smaller than the observed value for YPtBi observed in STM \cite{Baek2015} and point contact measurements (see below).

\subsection{Soft point-contact spectroscopy}

\begin{figure}
\includegraphics[width=0.7\linewidth]{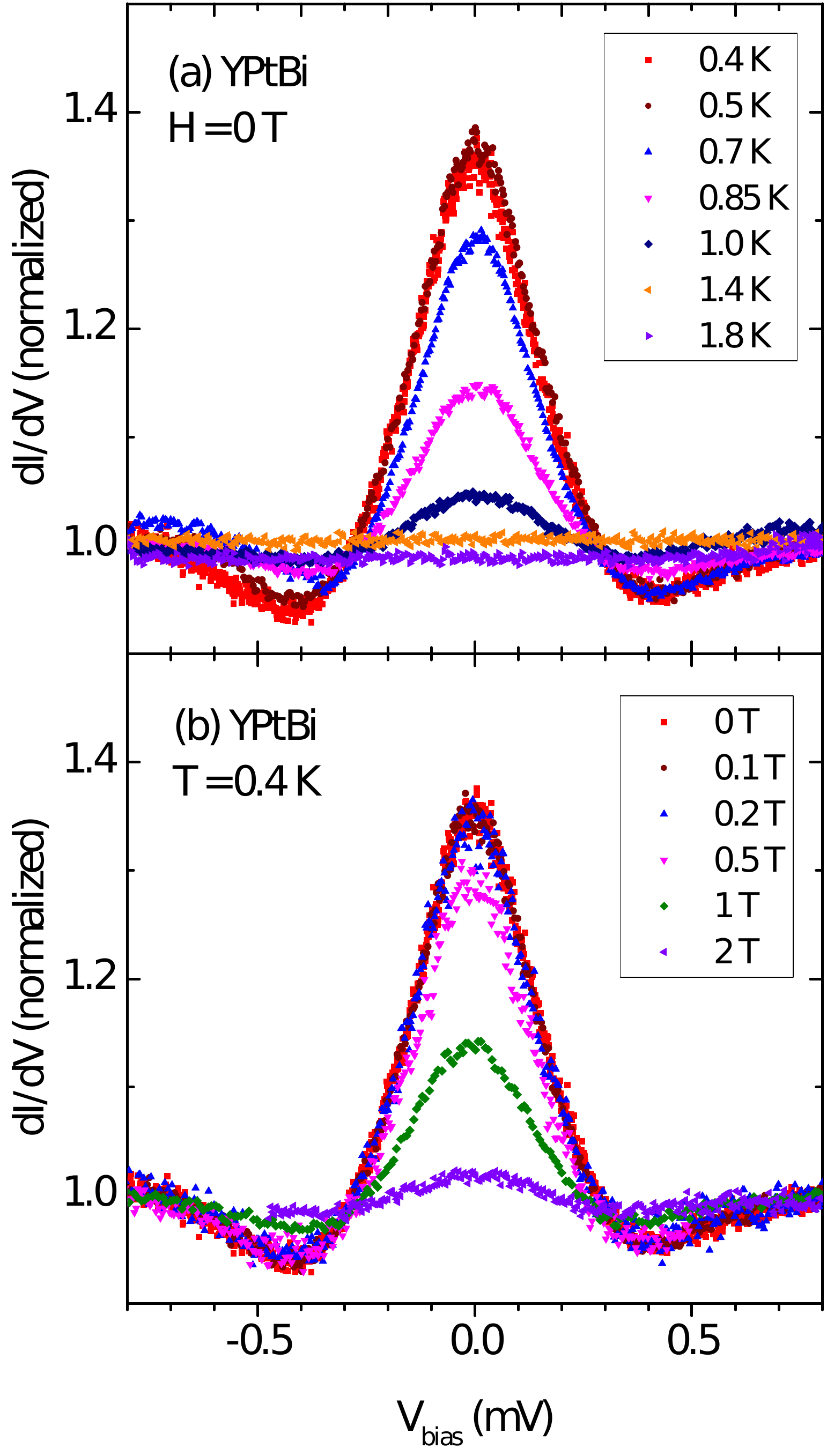}%
\caption{\label{fig:pcs} Soft point-contact spectroscopy conductance spectra of YPtBi as a function of (a) temperature and (b) magnetic field. An enhancement of roughly 40\% is observed at zero bias for 0.4 K base temperature in zero field. Note the persistence of the conductance peak associated with the superconducting gap to temperatures above $T_c = 0.78$~K, which may have a relation to the associated theory of $j=3/2$ pairing \cite{Brydon2016}.} 
\end{figure}

\begin{figure*}
\includegraphics[width=1\linewidth]{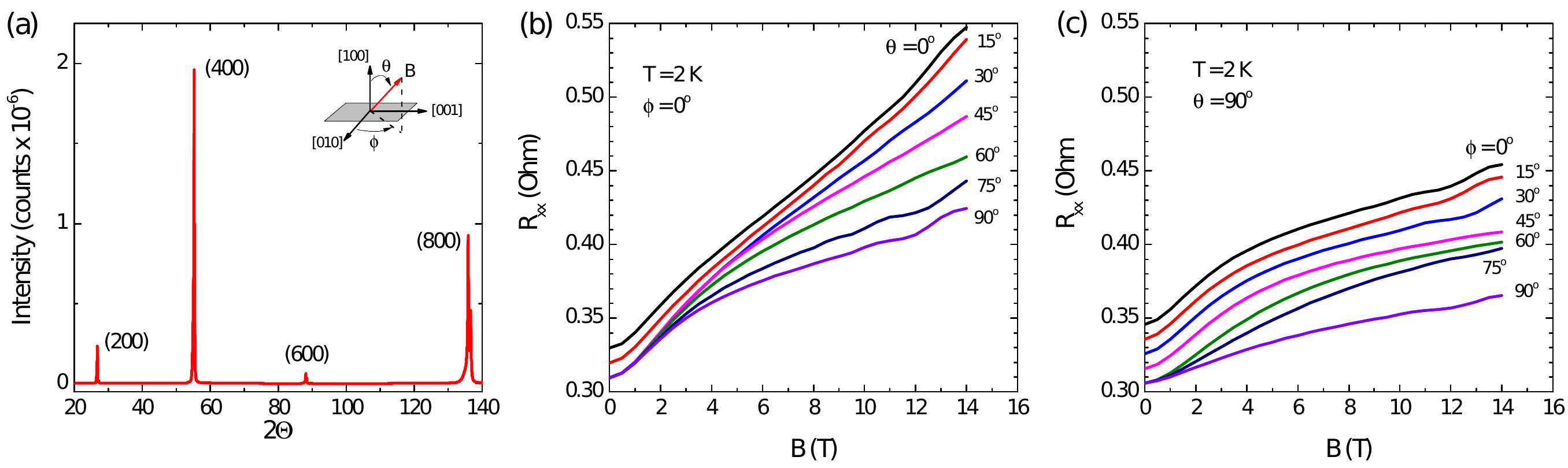}%
\caption{\label{fig:qo1} (a) Single crystal x-ray diffraction pattern of (100) surface of YPtBi used for angle-dependent quantum oscillation experiments. A schematic defines two controlled angles $\theta$ and $\phi$. Angle-dependent magnetoresistance (b) with varying $\theta$ at fixed $\phi=0$\degree~and (c) with varying $\phi$ at fixed $\theta=90$\degree. Shubnikov-de Haas quantum oscillations are evident in most of the orientations.
} 
\end{figure*}

Point contact spectroscopy was performed using the ``soft'' point contact technique, with conductance spectra measured on a polished surface of single crystal YPtBi. Contact junctions were prepared by attaching a bent 25~$\mu$m gold wire to a sample using DuPont 4929N silver paste. Typical junctions were on the order of 0.1~$\mu$m in diameter achieved by pre-coating the surface with a thin layer of stycast with a small hole. $dI/dV$ spectra were measured in a commercial $^3$He cryostat. AC conductivity vs$.$ DC bias voltage was measured in zero magnetic field at several temperatures between 400~mK and 1.8~K and at multiple applied magnetic fields at 400~mK. 

Figure \ref{fig:pcs} shows the results from point-contact spectroscopy measurements on a single crystal YPtBi. The spectra show strong temperature-and field-dependence on the temperature-and field-independent flat background conductance at high bias as shown respectively in Fig \ref{fig:pcs}(a) and (b). Thus, the observed shape of spectra represents the superconducting gap. The data in Fig$.$ \ref{fig:pcs} were normalized by dividing by the high bias value for $dI/dV$. It is notable that $dI/dV$ increases monotonically up to $V_{bias}$ = 0. In addition to the shape of the curves, other aspects of the data are also unusual compared to what is expected for conventional superconductivity. First, the zero bias enhancement shown in Fig$.$ \ref{fig:pcs} persists at values of temperature and magnetic field that are above the values of $T_c$ and $H_{c2}$ that were determined from the resistivity measurements \cite{Butch2011,Bay2012}. Second, the size of these gap features is also quite large compared to the expected BCS gap value of $\Delta_0$ = 1.76${k_BT_c}$ = 0.1~meV for a $T_c$ of 0.7 K. In fact, based on the peak width the gap size appears to be at least twice the BCS value. This large deviation from the BCS theory provides further evidence against the conventional $s$-wave model.

\subsection{Angle-dependent magnetoresistance}

Angle-dependent magnetoresistance was measured to study the spin-split Fermi surface of YPtBi. A sample was cut out of (100) plane confirmed by a single-crystal diffraction pattern as shown in Fig. \ref{fig:qo1}(a). The angle-dependence of the longitudinal magnetoresistance $R_{xx}$ was measured by using a single-axis rotator at various orientations with two controlled angles $\theta$ and $\phi$ defined in Fig. \ref{fig:qo1}(a). Panel (b) shows angle-dependent magnetoresistance data with varying $\theta$ from 0\degree~to 90\degree~at $\phi=0$\degree. At most of the angles, Shubnikov-de Haas (SdH) quantum oscillations are visible on smoothly increasing magnetoresistance. Similar experiments with varying $\phi$ from 0\degree~to 90\degree~at $\theta=90$\degree~were done and presented in Fig. \ref{fig:qo1}(c). In both configurations, SdH oscillations show angle-dependence where the amplitude and phase changes at different orientations of the applied magnetic field implying the SdH consists of multiple frequencies with varying frequencies and relative phases. The oscillations at 0\degree~and 90\degree~in both configurations appear the same, confirming the assignment of crystallographic orientation in the schematic.

\begin{figure}
\includegraphics[width=1\linewidth]{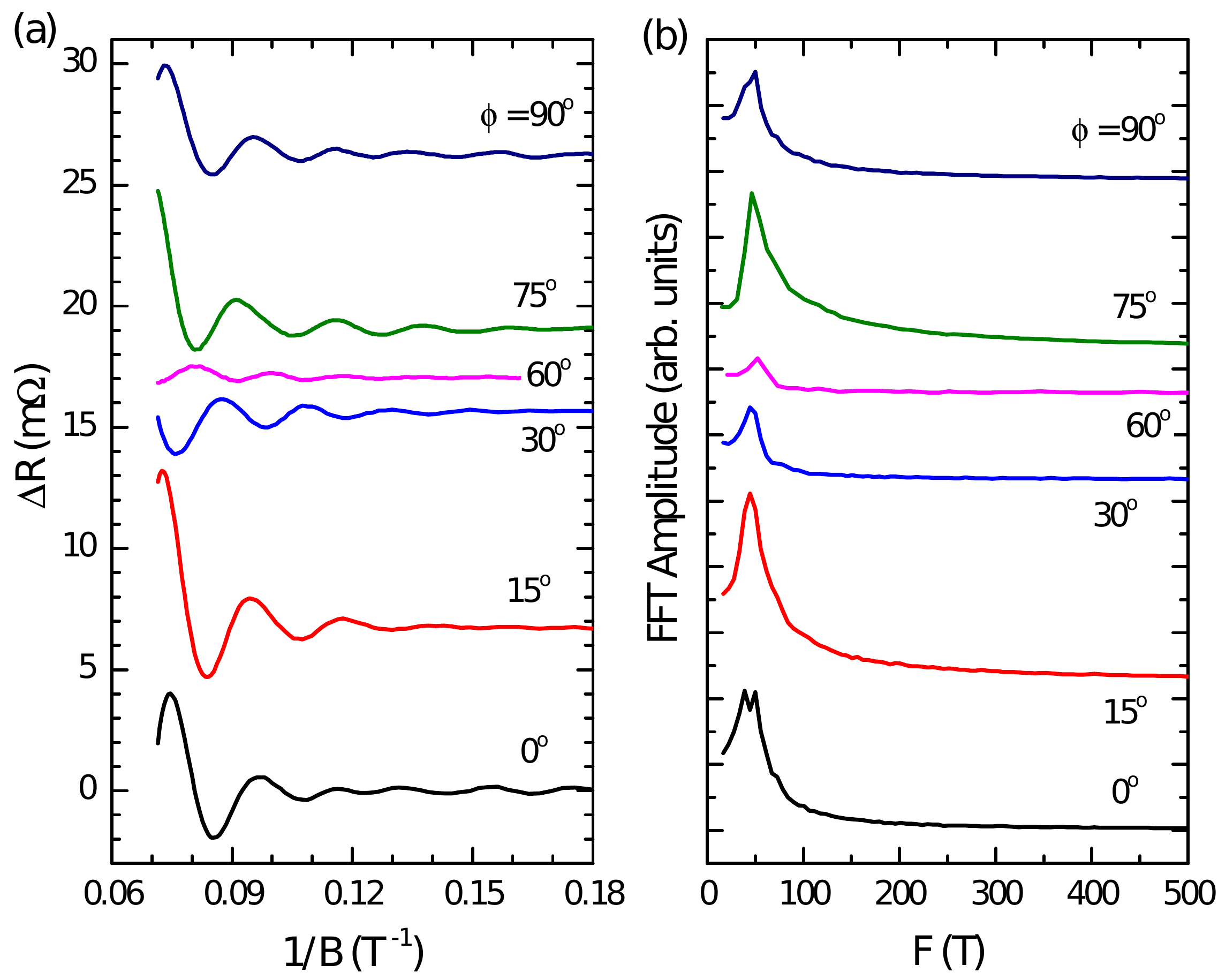}%
\caption{\label{fig:qo2} (a) SdH quantum oscillations extracted from the magnetoresistance at selected $\phi$ angles with fixed $\theta=90$\degree defined in Fig. \ref{fig:qo1}(a). (b) Fast Fourier transform spectra obtained from the quantum oscillations shown in the panel (a).
} 
\end{figure}

Figure \ref{fig:qo2}(a) shows the SdH quantum oscillations obtained from the magnetoresistance data in Fig. \ref{fig:qo1}(c). The angle-dependence of SdH oscillation is clear while the oscillations at $\phi=0$\degree~and 90\degree~oscillations are nearly identical in both of which display a beating node around $B^{-1}=0.12$ T$^{-1}$. We employ the fast Fourier transform (FFT) to determined the frequency at each angle. The FFT-spectra are shown in Fig. \ref{fig:qo2}(b). At first glance, the spectrum shows a peak around $F= 46$ T at all angles. The angle-dependent frequency is shown in Fig. 3(b) in the main text. In some orientations, it shows a broad feature with double peaks, and the two peaks are clearly resolved in the data for $\phi=0$\degree~with two frequencies of $39\pm3$ T and $50\pm3$ T where the error bar is the frequency resolution of FFT. We attributed the observation of two frequencies to existence of spin-split Fermi surfaces. Using a theory by Mineev and Samokhin\cite{Mineev2005}, the estimated energy of spin-orbit coupling is about 1 meV, assuming the Zeeman interaction is much smaller than the spin-orbit interaction.

\begin{figure}
\includegraphics[width=1\linewidth]{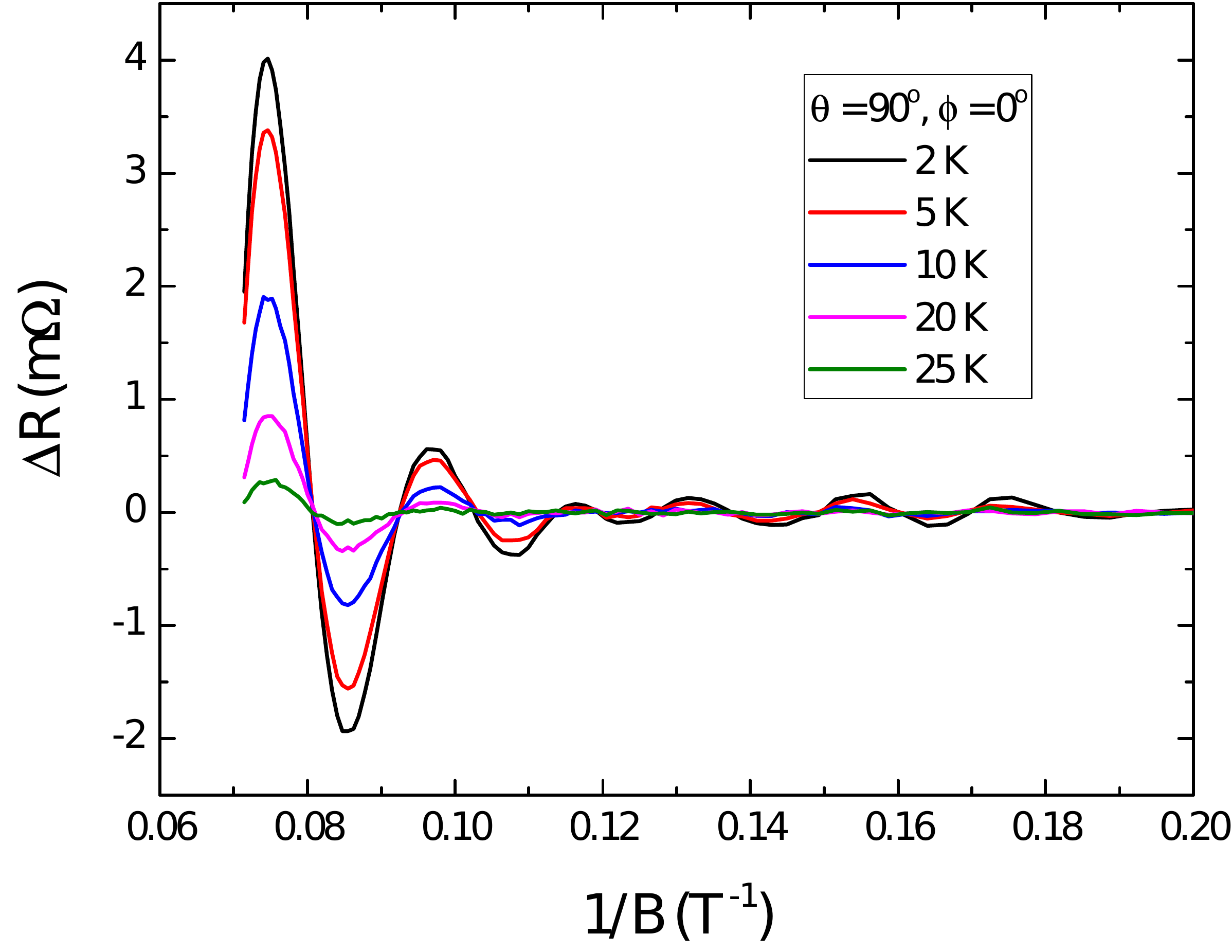}%
\caption{\label{fig:qo3} Temperature-dependent quantum oscillations with $\theta=90$\degree~and $\phi=0$\degree defined in Fig. \ref{fig:qo1}(a).} 
\end{figure}

To learn the nature of the Fermi surfaces associated with two resolved frequencies, we measured temperature-dependent SdH quantum oscillations with field along [010], i.e., $\theta=90$\degree~and $\phi=0$\degree. Temperature dependent SdH oscillations are displayed in Fig. \ref{fig:qo3}, FFT-spectra of which are presented in Fig. 3(c) of the main text. We determined effective mass $m^*=0.11m_e$ from temperature dependence of the amplitude by using the Lifshitz-Kosevich theory (see Fig. 3(d) in the main text). They have nearly identical effective mass as expected for spin-split Fermi surfaces.

\bibliography{YPtBi}

\end{document}